\documentclass[sigconf]{acmart}
\usepackage{graphicx}
\usepackage{listings}
\usepackage{bchart}
\usepackage{tikz}
\usepackage{pgf-pie}
\AtBeginDocument{%
  \providecommand\BibTeX{{%
    \normalfont B\kern-0.5em{\scshape i\kern-0.25em b}\kern-0.8em\TeX}}}

\copyrightyear{2021}
\acmYear{2021}
\setcopyright{ccby}
\acmConference[K-CAP '21] {Proceedings of the 11th Knowledge Capture Conference}{December 2--3, 2021}{Virtual Event, USA.}
\acmBooktitle{Proceedings of the 11th Knowledge Capture Conference (K-CAP '21), December 2--3, 2021, Virtual Event, USA}
\acmPrice{}
\acmISBN{978-1-4503-8457-5/21/12}
\acmDOI{10.1145/3460210.3493567}

\begin{document}
\fancyhead{}

\title{Living Literature Reviews}

\author{Michel Wijkstra}
\authornote{Both authors contributed equally to this research.}
\email{m.wijkstra@vu.nl}
\orcid{0000-0002-2363-5880}
\affiliation{%
  \institution{Vrije Universiteit Amsterdam}
  \streetaddress{De Boelelaan 1085}
  \country{The Netherlands}
  \postcode{1081 HV}
}

\author{Timo Lek}
\authornotemark[1]
\email{t.m.lek@vu.nl}
\orcid{0000-0002-3429-2879}
\affiliation{%
  \institution{Vrije Universiteit Amsterdam}
  \streetaddress{De Boelelaan 1085}
  \country{The Netherlands}
  \postcode{1081 HV}
}

\author{Tobias Kuhn}
\email{t.kuhn@vu.nl}
\orcid{0000-0002-1267-0234}
\affiliation{%
  \institution{Vrije Universiteit Amsterdam}
  \streetaddress{De Boelelaan 1085}
  \country{The Netherlands}
  \postcode{1081 HV}
}

\author{Kasper Welbers}
\email{k.welbers@vu.nl}
\orcid{0000-0003-2929-3815}
\affiliation{%
  \institution{Vrije Universiteit Amsterdam}
  \streetaddress{De Boelelaan 1085}
  \country{The Netherlands}
  \postcode{1081 HV}
}

\author{Mickey Steijaert}
\email{mickey.steijaert@gmail.com}
\orcid{0000-0003-1460-8757}
\affiliation{%
  \institution{Vrije Universiteit Amsterdam}
  \streetaddress{De Boelelaan 1085}
  \country{The Netherlands}
  \postcode{1081 HV}
}

\renewcommand{\shortauthors}{Wijkstra and Lek, et al.}

\begin{abstract}
Literature reviews have long played a fundamental role in synthesizing the current state of a research field. However, in recent years, certain fields have evolved at such a rapid rate that literature reviews quickly lose their relevance as new work is published that renders them outdated. We should therefore rethink how to structure and publish such literature reviews with their highly valuable synthesized content. Here, we aim to determine if existing Linked Data technologies can be harnessed to prolong the relevance of literature reviews and whether researchers are comfortable with working with such a solution. We present here our approach of ``living literature reviews'' where the core information is represented as Linked Data which can be amended with new findings after the publication of the literature review. We present a prototype implementation, which we use for a case study where we expose potential users to a concrete literature review modeled with our approach.
We observe that our model is technically feasible and is received well by researchers, with our ``living'' versions scoring higher than their traditional counterparts in our user study.
In conclusion, we find that there are strong benefits to using a Linked Data solution to extend the effective lifetime of a literature review. 

\end{abstract}




\maketitle

\section{Introduction} 

Literature reviews are an invaluable asset in all branches of science for researchers to get an overview of studies relevant to their research questions. In communication science, for example, often a wide range of empirical approaches are applied to the same question, reflecting the variety in backgrounds of communication scholars (e.g., linguistics, political science, psychology) \cite{mcquail2010mcquail} and literature reviews are critical to bringing these different
approaches together. Such reviews, however, require a lot of work to be compiled and are quickly outdated, in particular in fast-moving fields such as social media research. Moreover, these reviews often present valuable overviews and aggregations of the field (e.g., number of studies supporting a hypothesis) but this data cannot easily be accessed or integrated due to the traditional paper format.

This research proposes the concept of \emph{living literature reviews}, with which literature reviews would be machine-interpretable, interoperable, and automatically updated. To achieve this, we apply the concept and technology of nanopublications \cite{kuhn2013broadening}. Nanopublications are a container format to represent and publish scientific (and other kinds of) statements as small pieces of Linked Data \cite{groth2010anatomy}. They come with formal provenance and metadata attached, and can be published and queried in a reliable and redundant manner through the existing decentralized server network \cite{kuhn2016decentralized}.

By using nanopublications to represent the information that is captured in a literature review, we can greatly extend the usefulness and value of this information. This is because it allows us to extend, compare, and contrast the available information continuously as new literature is published. In the long term, this type of system could allow us to analyze the metadata of the entire catalog of literature stored in the system, enabling us to discover connections or information that was previously hidden.

We present a study below to assess the feasibility and usability of our approach of using nanopublications to automatically update literature reviews. As the technical properties of the nanopublication landscape have been fairly well studied, we will here focus on the user aspects, i.e. the role of researchers consuming such literature reviews. Our research question is therefore whether nanopublications can yield us machine-interpretable, interoperable, and easily updatable literature reviews in an inclusive and provenance-aware manner. We designed a model to facilitate the information stored in a literature review, implement that model with a specific chosen case in communication science, create a prototype to assess the effectiveness, and explore additional possibilities created by the new system. We show that the prototype was positively evaluated by 22 people who are active in academia.



\section{Background}
Here we briefly discuss the relevant background about Linked Data and nanopublications, literature studies and their history, and how these topics have been combined in existing work.

\subsection{Linked Data and Nanopublictions} 
Linked Data can be explained as connected data on the web that comes from multiple sources and is structured in such a way that it is also machine-readable \cite{bizer2011linked}. The main building blocks to publish Linked Data are the Web Ontology Language (OWL) and the Resource Description Framework (RDF). The data itself is structured in so-called triples, consisting of a subject, a predicate, and an object. The properties and characteristics of the things mentioned in such triples are typically defined in ontologies. A wide range of such ontologies exist that define predicates and classes with which statements on these subjects can be formally expressed \cite{bizer2011linked,mcguinness2004owl}. Nanopublications can be seen as an RDF-based container format for representing and publishing Linked Data in a granular and provenance-aware manner \cite{kuhn2013broadening}. Nanopublications consist of three RDF graphs: assertion, provenance, and publication information. The assertion carries the core content of a nanopublication, for example stating in a formal way that a specific gene is related to a given disease. The provenance part of the publications describes where the information of this assertion originates from, for example by linking to a study or paper. Lastly, the publication information is the information about the publication itself such as by whom and when it was created \cite{groth2010anatomy}. Nanopublications can easily be created and published using the Java library for nanopublications and its command-line tool \cite{kuhn2015nanopub}. The identifiers that are used can be made immutable and verifiable by using trusty URIs. These trusty URIs contain a hash based on the content of the nanopublication, which makes them unique \cite{kuhn2014trusty}. Nanopublications can be published to a decentralized server network, which redundantly stores the nanopublications, which makes the network robust against server failures \cite{kuhn2016decentralized}. It is also possible to create nanopublication indexes to link to other nanopublications and thereby creating sets of nanopublications. In this way, entire datasets of nanopublications can be reliably referenced and versioned \cite{kuhn2017reliable}.

\subsection{Literature Reviews in Communication Science}
Literature reviews have become more important in the social sciences in the last twenty-five years. With the rise of computer technology and the growing amounts of research, both the opportunity and the need for synthesizing research grew \cite{cooper1998synthesizing}.

However, synthesizing research in the social sciences is challenging due to the strong interdisciplinary focus. The same questions in a field such as communication science are typically studied from various angles, by researchers from different disciplines that are often unfamiliar with each other's work. Herbst \cite{herbs2008} describes the field of communication science as postdisciplinary, recalling Menand's \cite{menand} interpretation that in a postdisciplinary field scholars do not work within fields but at intersections of materials and theories. According to Herbst, this poses problems for literature reviews. Specifically, she mentions that the ``wheel reinvention problem'', where studies from the now vast landscape of interdisciplinary literature may be overlooked, causes missed opportunities that might have altered new work in a fundamental way.     

As a result, literature reviews are in high demand, but best practices and formats for how to compile them still need to be established. Burgers et al. \cite{BURGERS2019102} make a similar observation, and propose ways to make literature reviews more suitable for interdisciplinary topics. Their work explains multiple ways to conduct reviews that can be effectively utilized in research collaborations. This will potentially increase their efficacy and popularity in the future.

\subsection{Linked Data and Literature Reviews}

Several ontologies are available that can help to represent literature reviews as Linked Data. A suite of such ontologies is grouped together in the SPAR ontologies \cite{peroni2018spar}. One of these ontologies is the FaBiO ontology describing publishable entities such as papers or journals. Together with the CiTO ontology, which is also part of the SPAR ontologies, different kinds of relations between papers can be described, for example that a paper reviews another paper. In order to keep track of the origin of the information, the PROV-O ontology is particularly useful and has been applied extensively in related approaches \cite{lebo2013prov}. 
The cooperation databank
embodies an interesting approach to describe scientific studies and their results in a semantic manner \cite{spadaro2020cooperation}, applying an ontology that includes features like study size and land of focus. Initiatives like the Open Research Knowledge Graph (ORKG) \cite{Jaradeh} and the Artificial Intelligence Knowledge Graph (AI-KG) \cite{oro71736} have previously extracted and described research using similar methods.
Lastly, there are existing approaches that allow us to talk about scientific statements, such as claims or hypotheses, with Linked Data terms. One of these approaches are AIDA sentences \cite{kuhn2013broadening}. These are English sentences that are Atomic, Independent, Declarative, and Absolute (AIDA) sentences. The core idea behind AIDA sentences is that they can be used as identifiers for the statements they represent. As such, they can help us to link these statements among each other and to other relevant entities, such as researchers and studies. In RDF, they are represented as URLs that include the complete AIDA sentence in URL encoding \cite{tobias2018using}.
To establish links between AIDA sentences, the HYCL ontology \cite{kuhn2017hycl} can be used, for example to state that a statement is more specific than another statement.



\section{Approach} 

\begin{table}[]
\centering\small
\caption{Prefixes and URIs of used ontologies}
\begin{tabular}{l|l}
\textbf{Prefix} & \textbf{Ontology URI} \\ \hline
cdoc & https://data.cooperationdatabank.org/vocab/class/ \\
cdop & https://data.cooperationdatabank.org/vocab/prop/ \\
foaf & http://xmlns.com/foaf/ \\
cito & http://purl.org/spar/cito/ \\
dct & http://purl.org/dc/terms/ \\
fabio & http://purl.org/spar/fabio/ \\
hycl & http://purl.org/petapico/o/hycl\# \\
llr & https://w3id.org/livingreviews/vocab/ \\
prov & http://www.w3.org/ns/prov\# \\
\end{tabular}
\label{tab:ontologies}
\end{table}

Here, we introduce our general approach and model of living literature reviews and our prototype implementation, and we sketch how updates can be represented and published.

\subsection{Living Literature Reviews} 
The general idea behind living literature reviews is to represent the core content of such documents in Linked Data and to allow for updates in the same Linked Data format to be added after the document itself has been published. A piece of literature (including literature reviews) can be seen as a collection of statements made by the authors of the work.
In the current way of writing literature, these statements are so closely intertwined with the text that only humans are able to extract them. In our system, we rely on Linked Data identifiers, which are formally structured versions of the statements made by the authors. By attaching these formal identifiers to their informal counterparts in the publication we make this publication machine-interpretable and at the same time automatically updatable when new information is added.
Subsequently, this allows literature written in this format to be compared, aggregated, and reasoned upon.
Our approach is based on several components introduced above, including FaBiO to represent publication types, AIDA sentences to represent claims, HYCL to represent their relations, PROV-O to express provenance, and nanopublications to package and publish this information. Our model and prototype are introduced in more detail in the remainder of this section.
\begin{figure}[tb]
    \centering
    \includegraphics[width=0.45\textwidth]{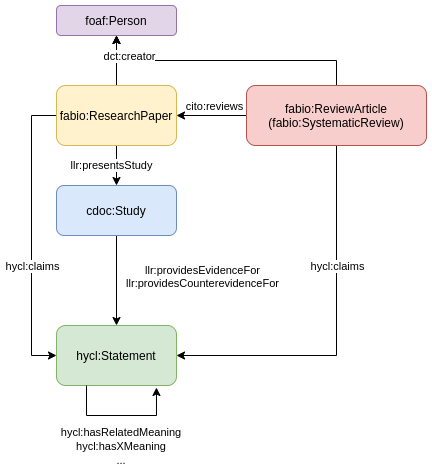}
    \caption{General scheme of how the data is linked. }
    \label{fig:basicstructure}
\end{figure}

\subsection{Model}

Figure \ref{fig:basicstructure} shows the schematic data structure of our model, including the links between the entities. The model centers on three core entities: review articles, research papers, and studies. Review articles review multiple research papers and these research papers can themselves describe multiple studies. Papers are identified by their DOI (if they have one) and authors by their ORCID identifiers, and these identifiers are used to link them. The statements in this model correspond to the core claims that are typically made in the conclusions of the research papers. The studies presented in these papers typically provide some sort of evidence for these statements. These statements can be linked to other statements with respect to their meaning, e.g. when one has a related, more general, or more specific meaning than another one, or when their meanings conflict with each other.

Next, we can have a look at how this model can be used by packaging instances thereof as nanopublications. Instances of the three main entities, that is review articles, research papers, and studies, are each described in their own nanopublication. Within these nanopublications, new identifiers are minted for the entities that do not yet have one, such as studies. For the research paper nanopublications, we are using the DOI service to automatically retrieve their bibliographic metadata in RDF format, which we then package as nanopublications. Review articles on top of that include information about which research papers they include. The nanopublications for the studies contain the metadata of each study, such as in which country the study was performed, which equipments or technologies were used, and what the size of the study was. On top of that, they contain the links to the statements these studies provide evidence for, by using AIDA sentences as described above. The statements themselves do not have a separate nanopublication, but are linked from other nanopublications with predicates such as ``hycl:claims'' or ``llr:providesEvidenceFor''. Separate nanopublications do exist for inter-relations of statements, which consist of an assertion with a single triple linking the statements. Table \ref{tab:ontologies} shows the full URIs for all used ontologies. We will show examples of such nanopublications in the next section.

Overall, our model connects the document level (papers) to the method level (studies) and down to the domain level (statements). Even though the actual statements are only identified and not fully represented in a formal way, this domain-level statement network can potentially be very valuable to find information about all the things that have already been said about a specific topic. Furthermore, the network of linked papers, studies, and other claims can tell something about the status and validity of claims. When there are two claims with a contradictory relations, it is possible to check by how many papers each claim is made and by how many different authors. When a claim is made by different authors and many different papers and the other claim is only made in one paper, then we can assume that it is more likely that the claim supported by many authors and papers is true.

\subsection{Viewer Prototype}\label{sec:prototype}

\begin{figure*}[tbp]
    \centering
    \includegraphics[width=1.0\textwidth]{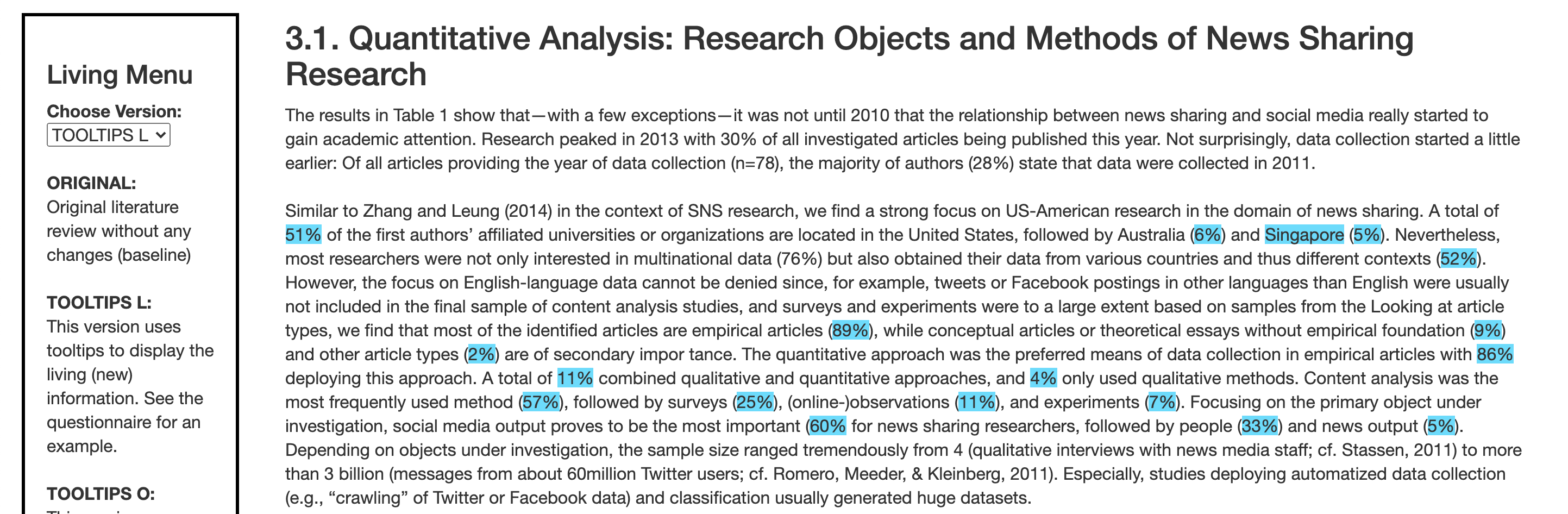}
    \caption{Screenshot of the viewer prototype. The drop-down box on the left enables users to change the version that is being displayed}
    \label{fig:prototype}
\end{figure*}

To demonstrate the capabilities of the described model an interactive viewer prototype had to be developed, a screenshot of which is shown in Figure \ref{fig:prototype}. This prototype displays the basic functionality of a living literature review. Using a manual process, a dataset was created by applying our living literature model on a chosen case (see below). This living literature RDF model was then made available through a Virtuoso triple store with a SPARQL endpoint. The user interface for the prototype is a simple website using HMTL5, JavaScript, and CSS made to look like a scientific publication. This website communicates with the SPARQL endpoint through a NodeJS abstraction layer.
The prototype displays multiple parts of an exemplary literature review which were deemed to be good examples to demonstrate our concept of living literature reviews. The prototype is capable of displaying information as it was originally written in a publication, as well as three different interpretations of what a living literature review could look like. These three ``living'' versions display variations in terms of the use of tooltips, highlights, and the mode in which text is displayed (e.g. original text in the tooltips versus living text in the tooltips). These different variants allowed us to evaluate user preference during the evaluation of the system.

\begin{figure}[tbp]
    \centering
    \includegraphics[width=0.35\textwidth]{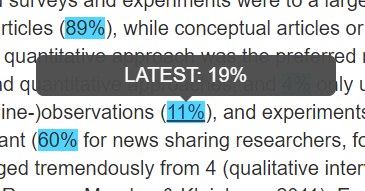}
    \caption{Example of a tooltip with blue highlighting }
    \label{fig:tooltip}
\end{figure}

The four different versions work as follows: Option 1 (Original) displays a publication as it was published, without any living elements. Option 2 (Tooltips L) shows the original publication. However, in this version, the elements that have living information available are highlighted and when the user moves the cursor across them a tooltip will show the living information, as shown in Figure \ref{fig:tooltip}. Option 3 (Tooltips O) provides similar functionality, but this version displays highlighted living information by default, with the original information appearing when the cursor is moved across it. Option 4 (Latest) is a fully living version of the document, displaying all the latest contributions registered in the system. This version does not contain any form of highlighting that indicates which elements are living. 

\subsection{Template for Adding Additions}
\begin{figure*}[t]
    \centering
    \includegraphics[width=1.0\textwidth]{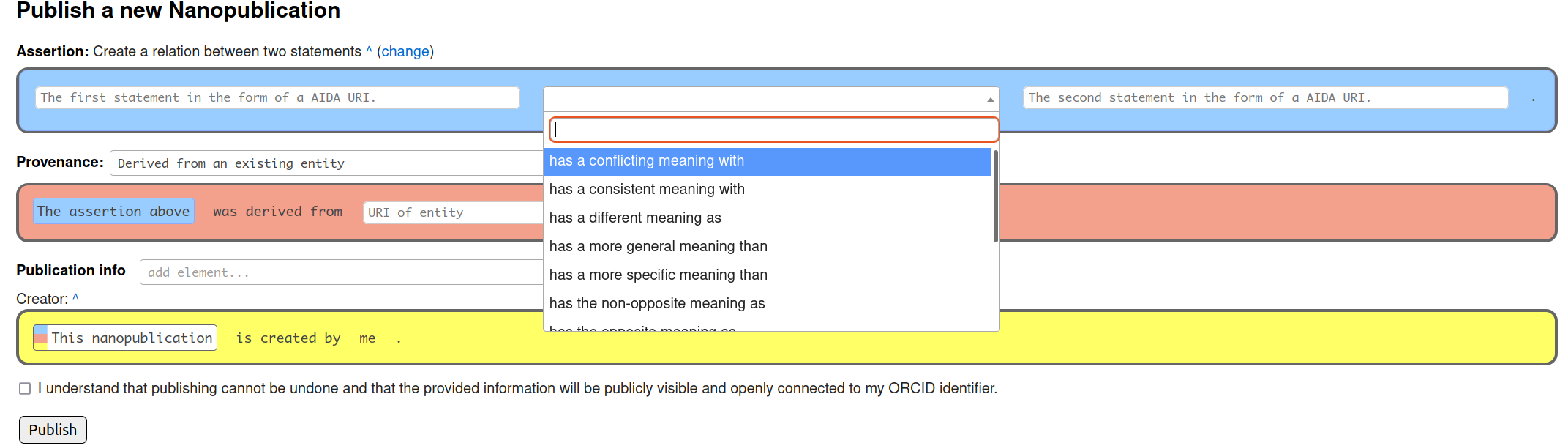}
    \caption{Screenshot of the Nanobench tool showing the template for publishing a relation between two statements}
    \label{fig:template}
\end{figure*}

To achieve our goal, it is not sufficient that the information that was provided in the paper is transformed in our model and nanopublications, as described above, but we also need to keep the data living, that is we need to be able to add new information easily when it comes available. We chose to use Nanobench \cite{kuhn2021semantic} to implement and showcase this aspect. Nanobench is a tool that can be downloaded and run locally to create and publish nanopublications based on templates in a decentralized way. To make our model living and updatable for everyone, templates can be created for each of the nanopublications that can be found in the model. In this way, a user can just fill in the template to create a new nanopublication that will be added to the web of nanopublications. New research papers, studies, statements, and the relation between statements can thereby be easily added to the already available data to truly keep the data living.
Figure \ref{fig:template} shows an example of a Nanobench template displayed as a form. In this specific example, a relation between two statements can be created.

\section{Case Study Design and Implementation} 
In this section, we explain how we selected a specific literature review for our case study. We elaborate on the data collection process and show examples of the created nanopublications.

\subsection{Paper Selection \&  Data Collection}

For this case study, we wanted to cover a recent literature review of a manageable but non-trivial size in a fast-moving research field, and with access to additional information that became available after the original publication date.
For these reasons, we selected a literature review in the area of social media studies, namely the publication by K\"umpel et al. called ``News Sharing in Social Media: A Review of Current Research on News Sharing Users'' \cite{Kumpel}. This paper consists of several tables with metadata information and several statements that can be made updatable and machine-interpretable with our living literature reviews approach. After contacting the authors they provided us with all the data that they collected for their review paper, including additional data points that were collected after the paper was finalized. We could then use this information to showcase our approach, create the nanopublications, and make a living version of it with the prototype introduced above.

\subsection{Examples of nanopublications}
Here we give some examples of nanopublications of the different types that were created specifically for this case study. We use TriG RDF notation to do this (informally to increase readability). These nanopublications are explained in detail to show how they together form a network of Linked Data. The generated nanopublications can be found on Github\footnote{\url{https://github.com/ucds-vu/living-reviews/}} including the code to generate them.
\begin{lstlisting}[basicstyle=\footnotesize\ttfamily,frame=single,breaklines=true]
sub:assertion {
  <https://doi.org/10.1177%2F2056305115610141>
    a fabio:ReviewArticle ;
    cito:reviews
      <http://doi.org/10.1016/j.chb.2011.10.002>,
      <http://doi.org/10.1016/j.chb.2014.03.006>,
      <http://doi.org/10.1016/j.chb.2014.08.009>,
      <http://doi.org/10.1080/08824096.2013.843165>,
      <http://doi.org/10.1080/1369118X.2011.554572>,
      ...
      <https://doi.org/10.1177/1077699013482906>,
      <https://doi.org/10.1177/1931243114546448>,
      <https://doi.org/10.1177/2056305115610141>,
      <https://doi.org/10.1207/s15506878jobem4903_3>,
      <https://doi.org/10.1287/isre.1100.0339> .
}

sub:provenance {
  sub:assertion prov:wasDerivedFrom
    <https://doi.org/10.1177%2F2056305115610141> .
}
\end{lstlisting}

The RDF code above shows the nanopublication of the literature review. This nanopublication describes which papers the review paper reviews by means of linking to the DOIs of the reviewed papers. We minted new URIs for the few cases where papers did not have a DOI.

We can move on to the nanopublications about the papers themselves:
\begin{lstlisting}[basicstyle=\footnotesize\ttfamily,frame=single,breaklines=true]
sub:assertion {
  <http://doi.org/10.1109/HICSS.2010.412>
    a fabio:ResearchPaper;
    hycl:claims
      aida:'Altruistic motive is one of the main drivers of information sharing.',
      aida:'People share news to gain reputation, to draw people's attention, and to attain status among peers or other users.';
    cdop:study nanopubRAAySjE5:study,
      nanopubRA0cEZj_:study .
}

sub:provenance {
  sub:assertion prov:wasDerivedFrom
    <http://doi.org/10.1109/HICSS.2010.412> .
}
\end{lstlisting}
The subject in the assertion of such a nanopublication is the DOI of the paper and through this DOI the nanopublication is linked to other nanopublications in the model. Next, the nanopublication contains the claims that are made in the form of an AIDA sentence, represented as a URI. In order to make these AIDA URIs easier to read, we show them here in a partially decoded manner (which is a slight violation of the used TriG notation). Lastly, the paper is linked to the studies that are described in it. In some cases, a paper can consist of multiple separate studies as can be seen in this example. The URIs of these studies point to other nanopublications where these studies are described:

\begin{lstlisting}[basicstyle=\footnotesize\ttfamily,frame=single,breaklines=true]
sub:assertion {
  sub:study a cdoc:Study, llr:EmpiricalArticle,
      llr:QuantatitiveAnalysis, llr:Survey;
    cdop:country dbpedia:United_States;
    cdop:overall "417";
    llr:firstAuthorOrigin dbpedia:United_States;
    llr:landOfFocus dbpedia:United_States;
    llr:primaryObject "People";
    llr:providesEvidenceFor aida:'People who share news in social media tend to perceive themselves as opinion leaders.';
    llr:theoreticalApproach "Uses and gratifications" .
}

sub:provenance {
  sub:assertion prov:wasDerivedFrom
    <https://doi.org/10.1177/1931243114546448> .
}
\end{lstlisting}
In a nanopublication containing a study, as shown above, information about the performed study can be found. To get this information for this case study, we relied on the rich information of supplemental tables provided by the authors of our chosen literature review. We performed some small manual data augmentation, such as mapping to DBpedia identifiers. The study's metadata displayed here includes information about the country where the study was performed, the study's group size, and more. Moreover, these nanopublications also link the studies to the claims for which the study provided evidence or counter-evidence for. The claims are made in the paper itself, but the evidence to support those claims has to come from the study that was performed in the paper.

Lastly, nanopublications containing relations were created to link related claims:
\begin{lstlisting}[basicstyle=\footnotesize\ttfamily,frame=single,breaklines=true]
sub:assertion {
  aida:'People who share news in social media tend to perceive themselves as opinion leaders.'
    hycl:hasRelatedMeaning aida:'People who share news in social media tend to have more friends or followers.' .
}

sub:provenance {
  sub:assertion prov:wasDerivedFrom 
    <https://doi.org/10.1177/2056305115610141> .
}
\end{lstlisting}
This concludes our description of the design and implementation of our case study. We can now evaluate the obtained data.

\section{Case Study Evaluation}

The evaluation is separated into two parts. We explain below the design and results of our descriptive analysis and our user study.

\subsection{Descriptive Analysis}

In the descriptive analysis the different groups of nanopublications are analyzed in more detail. Moreover, SPARQL queries are written to get a better insight in how all the nanopublications are linked together. 


The case study led to the creation of 450 nanopublications. They can be divided in several groups, as follows:
\begin{itemize}
\item 1 nanopublication representing the literature review itself and all the literature it is reviewing
\item 118 nanopublication containing general publication metadata, harvested from doi.org.
\item 118 nanopublications about the papers containing links to studies, statements, and the DOI nanopublications.
\item 163 nanopublications containing study information.
\item 31 nanopublications describing the relations between the different statements.
\item 19 nanopublications which are used to define classes of nanopublications, relations between nanopublications, or properties of nanopublications

\end{itemize}

Based on this distribution it is clear that the number of nanopublications for the DOI and papers is a one-to-one mapping. Furthermore, comparing this number to the numbers of studies present, shows that while 48\% of the papers only describe a single study, a small majority of the papers cover additional studies. Lastly, the number of relations between the statements is relatively low. The main reason for this is that in this study only a single review is used and not multiple reviews on a single topic. If that were the case more relations would appear as there is more room for supporting or conflicting information.

Looking more into depth on the statements and their relations. It can be seen that 84\% of all relations between the different statements is a``hasRelatedMeaning'' relation. Next, about 9\% of all relations is a ``hasMoreSpecificMeaningThan'' relation . Lastly, 6\% of all relations is a ``hasConflictingMeaning'' relation. As mentioned before the reason for the fact that almost all relations are of type ``hasRelatedMeaning'' is that only a single review paper is used in this study. When statements are included that are derived from different papers with different levels of detail multiple relations are possible such as ``hasMoreSpecificMeaningThan'' or ``hasMoreGeneralMeaningThan''.

With the designed structure of nanopublications, the origin of the statements and other things can also be analyzed in more detail. The statements made in this review all originate from different papers. Using SPARQL queries, it can be seen that 63\% of all statements were based on evidence provided by papers for which the data was collected in the United States, and 63\% statements are based on studies of which the first author originates from the United States, showing that the authors are likely to perform their studies in their own country. Lastly, a query can show that 44\% of the statements were made based on studies that have a study group size greater than 1000 (considering only the cases where the study group size is known). With such queries, one can therefore easily get an indication of the strength of evidence for a statement. As a further demonstration of the usefulness of such queries, the listing below shows an example of a query that calculates the percentage of statements for which the evidence was provided by a survey. This query returns 44.44\% for our case:

\begin{lstlisting}[basicstyle=\footnotesize\ttfamily,frame=single,breaklines=true]
prefix cdoc: <https://data.cooperationdatabank.org/vocab/class/>
prefix llr: <https://w3id.org/livingreviews/vocab/>

SELECT (xsd:float(?cnt)/xsd:float((COUNT(?statement)))*100 AS ?percentage) WHERE {
    ?study llr:providesEvidenceFor ?statement. {
        SELECT (COUNT(?state_survey) AS ?cnt) WHERE {
            ?study llr:providesEvidenceFor ?state_survey;
             a llr:Survey.
        }
    }   
}
\end{lstlisting}
Overall, based on this analysis it could be seen that with this system using statements linked to papers and their study, together with the relations between statements can be very powerful. Furthermore, querying this system can enable access to relevant information about the review and its content that was previously unavailable. 










\subsection{User Study Design}
The user study aims to determine if and how communication scientists prefer to interact with living literature reviews. We achieve this by performing a comparative analysis in which we analyze the opinions of academic staff and students who have explored our viewer prototype. Participants were requested to read the original text of a literature review and then to explore our living versions using our viewer prototype described in Section \ref{sec:prototype}. We recorded their opinion of the system using a questionnaire and assessed their feeling about applying our approach from the perspective of a researcher getting an overview of a topic, a researcher citing a topic, and a researcher writing a paper in this format.

The respondents were presented with three questions regarding their satisfaction with each different version of the living literature viewer. These questions were scored using a seven-point Likert scale, with 1 being very unsatisfied and 7 being very satisfied.
We asked the following questions: 
\begin{enumerate}
  \item ``Assuming you are a researcher looking for a good background reference on the topic to CITE in an article, how satisfied would you be with the different versions?''
  \item ``Assuming you are a researcher reading this review to get an OVERVIEW of the topic, how satisfied would you be with the different versions?''
  \item ``Assuming you are an AUTHOR of a review paper, how satisfied would you be with it being published in the different versions?''
 \end{enumerate}
 
An important element of the proposed living literature system is deciding who should be allowed to make contributions to existing material. We proposed the following question to the respondents: ''Who do you think should be able to add updates, such as new papers, to an existing literature review?'' with the possible answer options of 1) only the original authors, 2) authors approved by the original authors, 3) researchers with verified credentials, or 4) anyone.

Additionally, respondents were requested to share what they liked about the system, what they did not like about the system and were free to share any additional feedback. We recruited participants via personal connections and internal research group mailing lists.

\subsection{User Study Results}
Over a period of two weeks, our survey accumulated 22 replies. Of these replies 8 (36.4\%) belonged to PhD candidates, 7 (31,8\%) belonged to assistant, associate, or full professors, 4 (18,2\%) belonged to Master students and 3 (13,6\%) belonged to postdocs. 45.5\% of the respondents were active in the field of communication science, 40.9\% were active in computer science and the remaining 12.5\% is shared across economics, information science, and business \& communications (4.5\% each).

\begin{figure}[tbp]
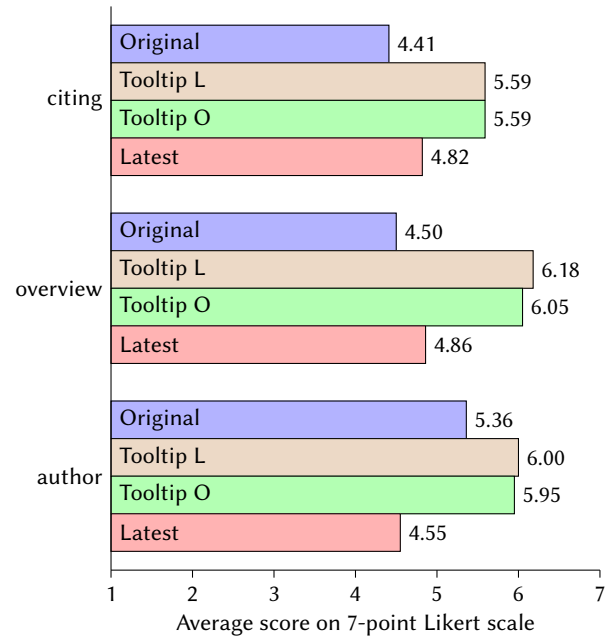

\centering
\begin{bchart}[max=7,step=1, min=1,width=6.5cm]
\bcbar[text=Original,color=blue!30]{4.41}
\bcbar[text=Tooltip L,color=brown!30]{5.59}
\bclabel{citing}
\bcbar[text=Tooltip O,color=green!30]{5.59}
\bcbar[text=Latest,color=red!30]{4.82}
\medskip
\bcbar[text=Original,color=blue!30]{4.50}
\bcbar[text=Tooltip L,color=brown!30]{6.18}
\bclabel{overview}
\bcbar[text=Tooltip O,color=green!30]{6.05}
\bcbar[text=Latest,color=red!30]{4.86}
\medskip
\bcbar[text=Original,color=blue!30]{5.36}
\bcbar[text=Tooltip L,color=brown!30]{6.00}
\bclabel{author}
\bcbar[text=Tooltip O,color=green!30]{5.95}
\bcbar[text=Latest,color=red!30]{4.55}
\bcxlabel{Average score on 7-point Likert scale}
\end{bchart}
\caption{Satisfaction with each version for citing (top), getting an overview (middle), and as the paper author (bottom)}
\label{fig:satisfaction}
\end{figure}

The results show in each case that the two interactive versions using tooltips and highlighting receive better scores than the versions showing only the original or only the latest version. 
The differences between the different roles (citing, overview, and author) are small, but we observe a small increase in preference for the living versions in the overview role, and a small increase in preference for the original version in the author role, when compared to the citing role.

When asked which version the respondents liked the most, 50\% chose the version with the latest information on the screen and the original information displayed in the tooltips (Tooltip O). 36\% picked the version with the original information on the screen and the latest information in the tooltips (Tooltip L). This shows that 86\% of respondents prefer one of the interactive versions, which is in line with the results we see in Figure \ref{fig:satisfaction} The remaining 14\% is shared between the latest information (9\%) and the original information (5\%).

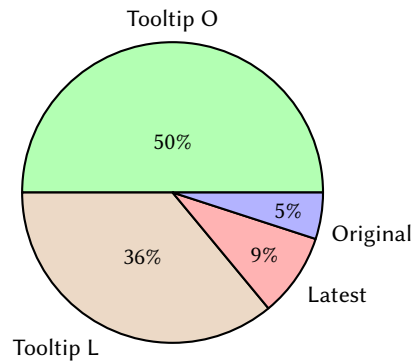
\begin{figure}
    \centering
    \sffamily
    \begin{tikzpicture}
\pie[pos={10,0},radius=2,color={green!30,brown!30,red!30,blue!30}]{
    50/Tooltip O,
    36/Tooltip L,
    9/Latest,
    5/Original
    }
\end{tikzpicture}
    \caption{Preferred versions of the respondents}
    \label{fig:Preferred}
\end{figure}

When asked who should be allowed to make contributions to existing material, the majority of the respondents chose either researchers with verified credentials or authors approved by the original authors (35.6\% each) followed by the original authors and anyone (13.6\% each).

To get a clearer picture of what the respondents liked and disliked about the system we allowed them to write an open answer stating their opinion. Positive reactions include fondness about keeping the information in the literature reviews useful and updated (15 of 21) while preserving the original basis of the work (6 of 21). Some examples include:

\begin{quote}
    \emph{It is very nice to have a living overview, to get insights on the evolutions in the field, and how initial findings might have changed.} --- Respondent 21
\end{quote}

\begin{quote}
    \emph{(I like) That the article information is not outdated and that both the original and the latest changes can be seen. This is an important step in order to have such "live reviews" as the information is constantly changing in the digital world. The idea is great!} --- Respondent 18
\end{quote}

The question asking for negative reactions received 20 replies. The replies include concerns about the integrity and tractability of the later contributed information (3 of 20) , the text becoming unclear or losing its meaning (4 of 20), complaints about the version which only shows the latest information (6 of 20)  and general complaints about the design of the prototype application (4 of 20). These are two examples of comment excerpts:

\begin{quote}
    \emph{It asks a lot of the reader in terms of trusting the system (are the new numbers correct? Were they peer-reviewed of checked properly? etc)} --- Respondent 1
\end{quote}

\begin{quote}
    \emph{I also find it too much work for original authors to permanently need to update their past paper, while at the same time unfair to them if many people add the updates minimizing over time the contribution of the original authors.} --- Respondent 7
\end{quote}

\section{Discussion and Conclusions}
Concluding, in this paper we have created a model, built a prototype, and presented a case study in communication science to represent literature reviews in a machine-interpretable manner and automatically updatable format. To evaluate our approach we performed a descriptive analysis and a user study.
We showed that our model and approach is feasible and that interesting insights can be generated by directly querying the model using semantic technologies such as SPARQL.

Next, the user study gave us some interesting insights about how to represent the updated information in literature reviews. Our study group used for this user study was of high quality with only having responses of people that are active in academia. Although some concerns were expressed about integrity and the maintenance and the additional workload for authors, the majority of the respondents have a positive attitude to our solution to represent updated information using tooltips in a literature review. 

Overall, it can be concluded that we succeeded with our approach to represent literature reviews in a machine-interpretable manner using nanopublications. Interesting questions can be answered with SPARQL queries. Furthermore, with the Nanobench templates it is also easy to add new information to the web of Linked Data which can be used in the living literature reviews. Our user study has shown that the way how living literature reviews are represented also has a positive effect on how the information is perceived by readers. Lastly, when more reviews are made `living' the model will only get more powerful due to the fact that more information is available, making living literature reviews even more interesting.

\section{Acknowledgements}

We wish to thank the VU Network Institute for providing us with the opportunity to research this promising concept. We also wish to thank Anna Sophia K\"umpel and her co-authors for allowing us to use the data they gathered for their work to support ours.

\bibliographystyle{ACM-Reference-Format}
\bibliography{references.bib}


\end{document}